\documentclass[12pt]{article}

\usepackage{amssymb}
\usepackage{amsmath}
\usepackage{amsfonts}
\usepackage[dvips]{graphicx}

\numberwithin{equation}{section} 
\newcommand{\be}{\begin{equation}}

\newcommand{\ee}{\end{equation}}
\newcommand{\bea}{\begin{eqnarray}}
\newcommand{\eea}{\end{eqnarray}}
\newcommand \tr {\hbox{Tr}}

\begin{document}

\title{On Weyl-covariant channels}
\author{M. Fukuda\\
Statistical Laboratory, CMS,
University of Cambridge \\A. S. Holevo \\
Steklov Mathematical Institute} \maketitle

\abstract{ Formalism of discrete noncommutative Fourier transform is developed and
applied to the study of Weyl-covariant channels. We then extend a result in
\cite{KMNR} concerning a bound of the maximal output $2$-norm of a Weyl-covariant
channel. A class of channels which attain the bound is introduced, for which the
multiplicativity of the maximal output $2$-norm is proven. Complementary channels
are described which share the multiplicativity properties with the Weyl-covariant
channels.}

\section{A noncommutative Fourier transform}

A state of finite quantum system is represented by a positive operator $\rho
$ of trace one (density operator) in a Hilbert space $\mathcal{H}$ of
dimensionality $d$. The set of density operators in $\mathcal{H}$ is denoted
$\mathfrak{S}(\mathcal{H})$. A channel $\Phi $ is a completely positive \
(CP) trace-preserving (TP) map of the algebra $\mathfrak{M}(\mathcal{H})$ of
all operators in $\mathcal{H}$. Although the TP condition is redundant in
the context of our results,
we shall impose it just for notational convenience.

The maximal output $p$-norm of $\Phi $ is defined as
\begin{align}
\nu_p(\Phi):=\sup_{\rho \in \mathfrak{S}(\mathcal{H})}\|\Phi(\rho)\|_p,
\end{align}
where $\|\;\;\|_p$ is the Schatten $p$-norm:
$\|\rho\|_p:=(\mathrm{tr}|\rho|^p)^{\frac{1}{p}}$.

The current multiplicativity conjecture is that
\begin{align}  \label{mult}
\nu_p(\Phi \otimes \Omega)= \nu_p(\Phi)\nu_p(\Omega),
\end{align}
for arbitrary channels $\Phi$ and $\Omega$, and for $p \in [1,2]$. Note that the
inequality $\nu_p(\Phi \otimes \Omega) \geq \nu_p(\Phi)\nu_p(\Omega)$ is
straightforward. In this paper we consider the case $p=2$, which is still an open
problem  (see \cite{GLR},\cite{KNR},\cite{KR2} for some general results in this direction).
 
Let us choose an orthonormal basis $\{e_{k};k=0,\dots,d-1\}$ in
$\mathcal{H}$. Consider the additive cyclic group $\mathbf{Z}_{d}$
and define an irreducible projective unitary representation of the
group $Z=\mathbf{Z}_{d}\oplus \mathbf{Z}_{d}$ in $\mathcal{H}$ as
\begin{equation*}
z=(x,y)\mapsto W_{z}=U^{x}V^{y},
\end{equation*}
where $x,y\in $ $\mathbf{Z}_{d},$ and $U$ and $V$ are the unitary
operators such that
\begin{equation*}
U|e_{k}\rangle=|e_{k+1(\mathrm{mod}d)}\rangle,\qquad
V|e_{k}\rangle=\exp \left( \frac{2\pi \mathrm{i} k}{d}\right)
|e_{k}\rangle.
\end{equation*}

The \textit{discrete Weyl operators} $W_{z}$ satisfy relations
similar to the canonical commutation relations for Weyl operators on
$Z=\mathbf{\ R^{s} }\oplus \mathbf{R^{s}}$ (see e. g. \cite{H}):
\begin{align}
W_{z}W_{z^{\prime }}& =\exp (\mathrm{i}\langle y,x^{\prime }\rangle
)W_{z+z^{\prime }};  \label{pr1} \\
W_{z}W_{z^{\prime }}& =\exp (\mathrm{i}(\langle x^{\prime },y\rangle
-\langle y^{\prime },x\rangle ))W_{z^{\prime }}W_{z};  \label{pr2}
\end{align}
where $\langle y,x\rangle :=2\pi yx/d$.

For future use we introduce the duality form on $Z$
\begin{equation*}
\langle z^{\prime },z\rangle :=\langle x^{\prime },x\rangle +\langle
y^{\prime },y\rangle,
\end{equation*}
and the symplectic form
\begin{equation*}
\langle z^{\prime },Jz\rangle := \langle x^{\prime },y\rangle
-\langle y^{\prime },x\rangle,
\end{equation*}
where $J(x,y):=(y,-x).$

Instead of the relation $W_{z}^{\ast }=W_{-z}$ for the usual Weyl
operators, we have
\begin{equation}
W_{z}^{\ast }=\exp (\mathrm{i}\langle y,x\rangle )W_{-z}.
\label{pr4}
\end{equation}
Moreover,
\begin{equation}
\tr W_{z}W_{z^{\prime }}^{\ast }=d\delta _{zz^{\prime }}.
\label{pr3}
\end{equation}
Consider $\mathfrak{M}(\mathcal{H})$ as a Hilbert space with the
Hilbert-Schmidt inner product. The Weyl operators form an orthogonal
basis in $\mathfrak{M}(\mathcal{H})$. Hence for all $X\in
\mathfrak{M}(\mathcal{H} ) $
\begin{equation*}
X=\sum_{z}f_{X}(z)W_{z},\qquad \text{where}\qquad
f_{X}(z)=\frac{1}{d} \tr XW_{z}^{\ast }.
\end{equation*}
The correspondence $X\leftrightarrow f_{X}(z)$ is a discrete analog
of the \textquotedblleft noncommutative Fourier
transform\textquotedblright , see \cite{H}. It has Parceval-type
properties
\begin{equation}
\tr X=df_{X}(0);\qquad \tr X^{\ast }X=d\sum_{z}|f_{X}(z)|^{2}.
\label{par}
\end{equation}
It follows that for a state $\rho \in \mathfrak{S}(\mathcal{H})$
\begin{equation}
f_{\rho }(0)=\frac{1}{d};\quad \left\vert f_{\rho }(z)\right\vert
\leq \frac{ 1}{d}  \label{cf1}
\end{equation}
and
\begin{equation*}
\sum_{z\neq 0}|f_{\rho }(z)|^{2}\leq \frac{d-1}{d^{2}}.
\end{equation*}
Moreover, $\rho $ is a pure state if and only if $\tr \rho ^{2}=1$,
which is equivalent to
\begin{equation}
\sum_{z\neq 0}|f_{\rho }(z)|^{2}=\frac{d-1}{d^{2}}.  \label{cf2}
\end{equation}

The relation (\ref{pr4}) implies
\begin{equation}
\overline{f_{X}(z)}=\exp (-\mathrm{i}\langle y,x\rangle )f_{X^{\ast
}}(-z). \label{adj}
\end{equation}

A necessary and sufficient condition for a Hermitian $X\in
\mathfrak{M}( \mathcal{H})\ $to be positive is: \textit{the
}$d^{2}\times d^{2}$\textit{\ -matrix}
\begin{equation*}
\left[ f_{X}(z^{\prime }-z)\exp (\mathrm{i}\langle y,x-x^{\prime
}\rangle ) \right] _{z,z^{\prime }\in Z}
\end{equation*}
\textit{is nonnegative definite.} The necessity follows from
\begin{equation*}
\sum_{z,z^{\prime }}\bar{c}_{z}c_{z^{\prime }}f_{X}(z^{\prime
}-z)\exp ( \mathrm{i}\langle y,x-x^{\prime }\rangle )=\frac{1}{d}\tr
X\left( \sum_{z}c_{z}W_{z}\right) ^{\ast }\left( \sum_{z^{\prime
}}c_{z^{\prime }}W_{z^{\prime }}\right) \geq 0.
\end{equation*}
The proof of sufficiency is similar to that for the case of the
\textquotedblleft noncommutative Fourier transform\textquotedblright
, see \cite{H}.

\section{Multiplicativity for the Weyl-covariant maps and channels}

A linear map $\Phi $  of $\mathfrak{M}(\mathcal{H})$ is
\textit{Weyl-covariant} if
\begin{equation*}
\Phi (W_{z}XW_{z}^{\ast })=W_{z}\Phi (X)W_{z}^{\ast }
\end{equation*}
for all $z\in Z$ and $X\in \mathfrak{M}(\mathcal{H})$. Inserting $
X=W_{z^{\prime }}$ we find that $\Phi (W_{z^{\prime }})$ satisfies
the same relation (\ref{pr2}) as $W_{z^{\prime }},$ hence $\Phi
(W_{z^{\prime }})W_{z^{\prime }}^{\ast }$ commute with all $W_{z}.$
Therefore
\begin{equation}
\Phi (W_{z})=\phi (z)W_{z},  \label{act}
\end{equation}
where $\phi (z);z\in Z,$ is a complex function. By making a
normalization, we can always assume that $\phi (0)=1.$ The class of
such maps we denote $\mathfrak{W}_{1}(\mathcal{H}).$ We shall also
use the notation $\Vert A\Vert _{2}=\sqrt{\tr A^{\ast }A}.$

Defining the Fourier transform
\begin{equation*}
p_{\gamma }=\frac{1}{d^{2}}\sum_{z\in Z}\phi (z)\exp (i\langle
\gamma ,z\rangle ),
\end{equation*}
we have
\begin{equation*}
\phi (z)=\sum_{\gamma \in Z}p_{\gamma }\exp (-\mathrm{i}\langle
\gamma ,z\rangle ),
\end{equation*}
for $\gamma =(\alpha ,\beta )\in Z.$ The relation (\ref{pr2})
implies that
\begin{equation}
\Phi (X)=\sum_{\gamma }p_{\gamma }W_{J\gamma }XW_{J\gamma }^{\ast }.
\label{ch1}
\end{equation}

There is a simple formula for composition of two Weyl-covariant maps
\begin{equation*}
(\Phi _{1}\circ \Phi _{2})(\rho )=(\Phi _{2}\circ \Phi _{1})(\rho
)=\sum_{\gamma }p_{\gamma }W_{J\gamma }\rho W_{J\gamma }^{\ast },
\end{equation*}
where $p_{\gamma }=p_{\gamma }^{(1)}\ast p_{\gamma }^{(2)}$ is the
convolution of functions $p_{\gamma }^{(1)},p_{\gamma }^{(2)},$
defining the maps $\Phi _{1},\Phi _{2},$ since the action of the
composition on the maps is given by (\ref{act}), where $\phi
(z)=\phi _{1}(z)\phi _{2}(z).$

The map $\Phi $ is channel if and only if  $\{p_{\gamma }\}$ is
probability distribution on $Z,$ and $\phi (z)$ -- its
characteristic function \cite{H1} . The relation (\ref{ch1}) is then
the Kraus representation.

Our principal estimate is:

\textbf{Theorem 1.} \textit{Let }$\Phi \in
\mathfrak{W}_{1}(\mathcal{H})$ \textit{\ and $\hat{\rho}$ -- an
operator in $\mathcal{H}\otimes \mathcal{K}$. Then}
\begin{equation}
\Vert (\Phi \otimes \mathrm{Id}_{\mathcal{K}})(\hat{\rho})\Vert
_{2}^{2}\leq \frac{1}{d}(1-\max_{z\neq 0}|\phi (z)|^{2})\Vert
\tr_{\mathcal{H}}\hat{\rho} \Vert _{2}^{2}+\max_{z\neq 0}|\phi
(z)|^{2}\Vert \hat{\rho}\Vert _{2}^{2}. \label{ineq}
\end{equation}

\textit{Proof}\textsl{.} Defining
$A_{z}=\frac{1}{d}\tr_{\mathcal{H}}\hat{ \rho}(W_{z}^{\ast }\otimes
I)$, we have
\begin{equation*}
\hat{\rho}=\sum_{z}W_{z}\otimes A_{z}.
\end{equation*}
Note that
\begin{align}
\Vert \hat{\rho}\Vert _{2}^{2}& =\tr\left( \sum_{z^{\prime
}}W_{z^{\prime }}^{\ast }\otimes A_{z^{\prime }}^{\ast }\right)
\left( \sum_{z}W_{z}\otimes
A_{z}\right) =d\sum_{z}\tr A_{z}^{\ast }A_{z};  \notag \\
\tr_{\mathcal{H}}\hat{\rho}& =\sum_{z}\tr W_{z}\otimes A_{z}=dA_{0};
\notag
\\
\Vert \tr_{\mathcal{H}}\hat{\rho}\Vert _{2}^{2}& =d^{2}\tr
A_{0}^{\ast }A_{0}.
\end{align}
Next, we have
\begin{align}
\Vert (\Phi \otimes \mathrm{Id}_{\mathcal{K}})(\hat{\rho})\Vert
_{2}^{2}& = \tr\left( \sum_{z^{\prime }}\phi (z^{\prime })^{\ast
}W_{z^{\prime }}^{\ast }\otimes A_{z^{\prime }}^{\ast }\right)
\left( \sum_{z}\phi (z)W_{z}\otimes
A_{z}\right)  \notag \\
& =d\sum_{z}|\phi (z)|^{2}\tr A_{z}^{\ast }A_{z}  \notag \\
& =d\left( \tr A_{0}^{\ast }A_{0}+\sum_{z\neq 0}|\phi (z)|^{2}\tr
A_{z}^{\ast }A_{z}\right)  \notag \\
& \leq d\left( \tr A_{0}^{\ast }A_{0}+\max_{z\neq 0}|\phi
(z)|^{2}\sum_{z\neq 0}\tr A_{z}^{\ast }A_{z}\right)  \label{ner} \\
& =d\left( \left( 1-\max_{z\neq 0}|\phi (z)|^{2}\right) \tr
A_{0}^{\ast }A_{0}+\max_{z\neq 0}|\phi (z)|^{2}\sum_{z}\tr
A_{z}^{\ast }A_{z}\right)
\notag \\
& =\frac{1}{d}\left( 1-\max_{z\neq 0}|\phi (z)|^{2}\right) \Vert
\tr_{ \mathcal{H}}\hat{\rho}\Vert _{2}^{2}+\max_{z\neq 0}|\phi
(z)|^{2}\Vert \hat{ \rho}\Vert _{2}^{2}.  \notag
\end{align}
QED

In the case of one dimensional $\mathcal{K}$ the bound (\ref{ineq})
implies the following inequality for channel $\Phi$ obtained in
proposition 9 of \cite{KMNR}:
\begin{equation*}
\tr\Phi (\rho )^{2}\leq \frac{1}{d}\left( 1+(d-1)\max_{z\neq 0}|\phi
(z)|^{2}\right) .
\end{equation*}
Moreover, this proposition states that, in the case $d=3$, the
equality is attained here for a special pure state $\rho $. This
observation can be substantially generalized (see theorem 3 below).

\textbf{Theorem 2.} \textit{Let} $\Phi \in
\mathfrak{W}_{1}(\mathcal{H})$ \textit{be such that }
\begin{equation}
|\phi (z)|\leq 1;\quad z\in Z,  \label{phi1}
\end{equation}
\textit{and }
\begin{equation}
\nu _{2}(\Phi )=\frac{1}{\sqrt{d}}\left( 1+(d-1)\max_{z\neq 0}|\phi
(z)|^{2}\right) ^{\frac{1}{2}},  \label{equ}
\end{equation}
\textit{then the multiplicativity of the maximal output}
$2$\textit{-norm holds for }$\Phi \otimes \Omega ,$ \textit{where}
$\Omega $ \textit{is an arbitrary CP map.}

\textit{Proof.} We have
\begin{align}
\Vert \tr_{\mathcal{H}}(\mathrm{Id}_{\mathcal{H}}\otimes \Omega
)(\hat{\rho} )\Vert _{2}& =\Vert \Omega
(\tr_{\mathcal{H}}\hat{\rho})\Vert _{2}\leq \nu
_{2}(\Omega )  \notag \\
\Vert (\mathrm{Id}_{\mathcal{H}}\otimes \Omega )(\hat{\rho})\Vert
_{2}& \leq \nu _{2}(\mathrm{Id}_{\mathcal{H}}\otimes \Omega )=\nu
_{2}(\Omega ),
\end{align}
where the last equality follows from \cite{AH}. Replacing
$\hat{\rho}$ by $( \mathrm{Id}_{\mathcal{H}}\otimes \Omega
)(\hat{\rho})$ in Theorem 1 and using (\ref{phi1}) gives
\begin{align}
\Vert (\Phi \otimes \Omega )(\hat{\rho})\Vert _{2}^{2}& \leq
\frac{1}{d} (1-\max_{z\neq 0}|\phi (z)|^{2})(\nu _{2}(\Omega
))^{2}+\max_{z\neq 0}|\phi
(z)|^{2}(\nu _{2}(\Omega ))^{2}  \notag \\
& =\frac{1}{d}\left( 1+(d-1)\max_{z\neq 0}|\phi (z)|^{2}\right) (\nu
_{2}(\Omega ))^{2}.
\end{align}
Therefore by (\ref{equ})
\begin{equation*}
\nu _{2}(\Phi \otimes \Omega )\leq \nu _{2}(\Phi )\nu _{2}(\Omega
).\qquad
\end{equation*}
QED

Define the set of optimizers of $|\phi (z)|$ for $z\neq 0$
\begin{equation*}
\mathcal{E}_{\max }:=\{z:z\neq 0,z=\arg \max_{z\neq 0}|\phi (z)|\}.
\end{equation*}

For a unit vector $|\psi \rangle \in \mathcal{H}$ consider the
subset of $Z$ defined as
\begin{equation}
\mathcal{G}_{\psi }:=\{z:|\psi \rangle \text{ is an eigenvector of
}W_{z}\}.
\end{equation}
By (\ref{pr1}) $\mathcal{G}_{\psi }$ is a subgroup of $Z$ and
$|\mathcal{G} _{\psi} |\leq d$ as we shall see from the proof of
theorem 3.

\textbf{Theorem 3.} \textit{Let }$d$ \textit{be arbitrary. A
necessary condition for the equality } (\ref{equ}) \textit{is
}$\left\vert \mathcal{E} _{\max }\right\vert \geq d-1.$ \textit{A
sufficient condition is that there is a subgroup }$\mathcal{G}_{\psi
}\subseteq Z$ \textit{such that }$ \left\vert \mathcal{G}_{\psi
}\right\vert =d$ \textit{and}
\begin{equation*}
\mathcal{G}_{\psi }\setminus \{0\}\subseteq \mathcal{E}_{\max }.
\end{equation*}

\textit{Proof.} If (\ref{equ}) holds then there exists a pure state
$\rho $ such that equality holds in (\ref{ner}) with $A_{z}=f_{\rho
}(z).$ This implies $\mathcal{N}:=\left\{ z:z\neq 0,f_{\rho }(z)\neq
0\right\} \subseteq \mathcal{E}_{\max }$. Hence the necessity
follows from (\ref{cf1}) and (\ref {cf2}).

Let $|\psi \rangle $ be a common eigenvector for the unitaries
$W_{z};z\in \mathcal{G}_{\psi },$ with eigenvalues $c_{z}$ of
modulus 1, and let us show first that $\left\vert \mathcal{G}_{\psi
}\right\vert \leq d$. If $ \left\vert \mathcal{G}_{\psi }\right\vert
\geq d$, then the operator
\begin{equation}
X=\frac{1}{d}\left( I+\sum_{z\in \mathcal{L}}\bar{c}_{z}W_{z}\right)
, \label{deco}
\end{equation}
where $\mathcal{L\ }$is any subset of $\mathcal{G}_{\psi }\setminus
\{0\},$ such that $\left\vert \mathcal{L}\right\vert =d-1,$
satisfies $X|\psi \rangle =|\psi \rangle ,$ and $\tr X^{\ast }X=1$
by (\ref{par}). This can be only the case if $X=\rho _{0}=|\psi
\rangle \langle \psi |$. Then it follows: 1) $\left\vert
\mathcal{G}_{\psi }\right\vert =d,$ for otherwise the operator $\rho
_{0}$ would have several different decompositions (\ref {deco})
corresponding to different subsets $\mathcal{L};$ 2) under the
assumptions of the theorem
\begin{align}
\tr\Phi (\rho _{0})^{\ast }\Phi (\rho _{0})& =d\left(
\frac{1}{d^{2}} +\sum_{z\in \mathcal{G}_{\psi }\setminus \{0\}}|\phi
(z)|^{2}\left\vert
\frac{c_{z}}{d}\right\vert ^{2}\right)  \notag \\
& =\frac{1}{d}\left( 1+(d-1)\max_{z\neq 0}|\phi (z)|^{2}\right)
.\qquad \text{QED}
\end{align}

A subset $\mathcal{F}\subseteq Z=\mathbf{Z}_{d}\oplus \mathbf{Z}
_{d}$ will be called \textit{degenerate} if the symplectic form
vanishes on $\mathcal{F} :$
\begin{equation*}
\langle z^{\prime },Jz\rangle =0,\quad z^{\prime },z\in \mathcal{F}.
\end{equation*}
A subgroup of $Z$ generated by $\mathcal{F}$ is again a degenerate
subset. Let $\mathcal{F}$ be degenerate, then the operators
$W_{z};z\in \mathcal{F},$ all commute by (\ref{pr2}) and hence have
common eigenvector(s). We conclude that $\mathcal{F\subseteq
G}_{\psi }$ for some $\psi ,$ hence $\left\vert
\mathcal{F}\right\vert \leq d,$ and if the equality holds, then
$\mathcal{F}$ is a (maximal degenerate) subgroup of $Z.$

\textbf{Examples}

1) Consider the cyclic subgroup generated by an element $z\in Z$
\begin{equation*}
\mathcal{G}(z):=\{kz:k=0,1,\ldots d-1\}.
\end{equation*}
This subgroup is degenerate and $\left\vert
\mathcal{G}(z)\right\vert =d$ in the case where $z=(\alpha ,\beta )$
and $\alpha ,\beta ,d$ have no common nontrivial divisor, in
particular if $d$ is prime.

2) Assume $d=p_1p_2$, where $p_1,p_2$ are primes, then the subgroup
generated by two elements $(p_1,0)$ and $(0,p_2)$ is a maximal
degenerate noncyclic subgroup.

\textbf{Corollary. }If there is a maximal degenerate subgroup
$\mathcal{G} \subseteq \mathcal{E}_{\max }\cup 0,$ then (\ref{equ})
holds.

\textbf{Examples}

1) As noticed in \cite{KMNR}, the condition of Theorem 3 always
holds if $d=3 $ and $\Phi $ is a channel. By using the fact that
$2z_{0}=-z_{0}$ in case $ d=3$, our Theorem 2 implies the
multiplicativity of $2$-norm in case $|\phi (z)|=|\phi (-z)|,$ e.g.
the map $\Phi $ is hermitian.

2) Any unital qubit $(d=2)$ channel is unitarily equivalent to the
form
\begin{equation*}
\Phi (\rho )=\sum_{\gamma }p_{\gamma }\sigma _{\gamma }\rho \sigma
_{\gamma },
\end{equation*}
where $\gamma =0,x,y,z$ and $\sigma _{\gamma }$ are the Pauli
matrices (see e. g. \cite{KR}). But in the case $d=2$ the discrete
Weyl operators are
\begin{equation*}
W_{00}=I=\sigma _{0},\quad W_{01}=V=\sigma _{z},\quad
W_{10}=U=\sigma _{x},\quad W_{11}=UV=-i\sigma _{y}.
\end{equation*}
Thus any unital qubit channel is covariant with respect to the
projective representation of the group $\mathbf{Z}_{2}\oplus
\mathbf{Z}_{2}$ generated by these discrete Weyl operators.

For any $z\neq 0$ the cyclic group $\mathcal{G}(z)$ consists of two
elements $\{0,z\}$. Hence the assumption of the corollary is always
satisfied for the unital qubit channels. More generally, it holds
for arbitrary qubit map $ \Phi \in \mathfrak{W}_{1}(\mathcal{H}).$

3) The $d-$depolarizing channel
\begin{equation*}
\Phi (\rho )=\lambda \rho +(1-\lambda )\tr\rho \frac{1}{d}I
\end{equation*}
is unitarily covariant, hence Weyl-covariant. For this channel $\phi
(z)=\lambda $ for $z\neq 0,$ hence $\mathcal{E}_{\max }=Z\setminus
\{0\}$ and the assumption of the corollary is trivially satisfied.
Moreover, the conclusion holds for the map $\Phi $ with arbitrary
$\lambda \in \mathbf{C,} \left\vert \lambda \right\vert \leq 1.$

4) Let $\mathcal{G}$ be a subgroup of order $d$ and define a
function on $Z$ :
\begin{equation}  \label{phiab}
\phi (z)=
\begin{cases}
1, & z=0 \\
a+b, & z\in \mathcal{G}\setminus \{0\} \\
b, & z\notin \mathcal{G}
\end{cases}
,
\end{equation}
where $a,b$ are complex numbers to be restricted later. By the
Fourier transform,
\begin{align}
p_{\gamma }& =\frac{1}{d^{2}}\sum_{z}\phi (z)\exp (i\langle \gamma
,z\rangle
) \\
& =\frac{1}{d^{2}}
\begin{cases}
1+a(d-1)+b(d^{2}-1) & \gamma =0 \\
1+a(d-1)-b & \gamma \in \mathcal{G}^{\perp }\setminus \{0\} \\
1-a-b & \gamma \notin \mathcal{G}^{\perp }.
\end{cases}
\end{align}
Here $\mathcal{G}^{\perp }=\{z^{\prime }:\langle z^{\prime
},z\rangle =0,\forall z\in \mathcal{G}\}$ is a subgroup of $Z,$ and
we used
\begin{equation*}
\sum_{z\in \Gamma }\exp (i\langle \gamma ,z\rangle )=\left\{
\begin{array}{c}
\left\vert \Gamma \right\vert ,\quad \gamma \in \Gamma ^{\perp } \\
0,\quad \gamma \not\in \Gamma ^{\perp }
\end{array}
\right.
\end{equation*}
for a subgroup $\Gamma \subseteq Z.$ To see this, for each $\gamma
\not\in \Gamma ^{\perp }$ choose $\bar{z}\in Z$ such that $\langle
\gamma ,\bar{z} \rangle \neq 0$. Factor $\Gamma $ by
$\mathcal{G}(\bar{z})$ of the order, say, $N$. Then the sum over
each coset is $0$:
\begin{equation}
\sum_{k=0}^{N-1} \exp ({}\langle\gamma,z+k\bar{z}\rangle) =\exp({}
\langle\gamma,z\rangle) \sum_{k=0}^{N-1}
\exp({}\langle\gamma,\bar{z}\rangle k) =0.
\end{equation}

A direct calculation shows that the Weyl-covariant map $\Phi $
defined by the function (\ref{phiab}) can be written as
\begin{equation}
\Phi (\rho )=a\Psi (\rho )+b\rho +(1-a-b){\tr}\rho \frac{I}{d},
\label{ccom}
\end{equation}
where
\begin{equation}
\Psi (\rho )=\frac{1}{d}\sum_{z\in \mathcal{G}^{\perp }}W_{Jz}\rho
W_{Jz}^{\ast }.  \label{deph0}
\end{equation}
If the group $\mathcal{G}$ is maximal degenerate, then
$\mathcal{G}^{\perp }=J\mathcal{G}$. To see this, take $J(x^{\prime
},y^{\prime })=(y^{\prime },-x^{\prime })\in \mathcal{G}^{\perp }$.
This implies $\langle y^{\prime },x\rangle +\langle -x^{\prime
},y\rangle =0$ for all $(x,y)\in \mathcal{G}$ . Since $\mathcal{G}$
is maximal degenerate we have $\mathcal{G}^{\perp }\subseteq
J\mathcal{G}$ . The inverse inclusion is obvious. Thus (\ref
{deph0}) takes the form
\begin{equation}
\Psi (\rho )=\frac{1}{d}\sum_{z\in \mathcal{G}}W_{z}\rho W_{z}^{\ast
}. \label{deph}
\end{equation}

Assuming $\left\vert b\right\vert \leq 1,\left\vert a+b\right\vert
\leq 1$ gives the condition (\ref{phi1}). Moreover, if $a,b$ satisfy
the condition $ \left\vert a+b\right\vert \geq \left\vert
b\right\vert ,$ the map (\ref{ccom} ) has the property in the
corollary, giving another case for which the multiplicativity of
$2$-norm holds.

If $\mathcal{G}$ is maximal cyclic then $\Psi $ is a
\textquotedblleft completely dephasing channel\textquotedblright :
\begin{equation}
\Psi (\rho )=\sum\limits_{j=1}^{d}|h_{j}\rangle \langle h_{j}|\rho
|h_{j}\rangle \langle h_{j}|,  \label{phase}
\end{equation}
where $\left\{ h_{j}\right\} $ is the orthonormal basis of the
commuting operators $\left\{ W_{z};z\in \mathcal{G}\right\} $ as we
shall show in a moment. The sum is an expectation onto Abelian
subalgebra of operators diagonal in basis $\{|h_{k}\rangle \}$. In
this case the condition $ \left\vert a+b\right\vert \geq \left\vert
b\right\vert $ becomes redundant. In fact defining the Weyl
operators relative to the new basis $ \{|h_{k}\rangle \}$, we have
the relation (\ref{deph}), where $\mathcal{G} =\{k(0,1):k=0,\ldots
,d-1\}$. Then $\mathcal{G}^{\perp }=\{l(1,0):l=0,\ldots ,d-1\}$, and
\begin{align*}
|a+b|\geq |b|& \Rightarrow \mathcal{G}\setminus \{0\}\subseteq
\mathcal{E}
_{\max } \\
|a+b|\leq |b|& \Rightarrow \mathcal{G}^{\perp }\setminus
\{0\}\subseteq \mathcal{E}_{\max }
\end{align*}
so that the condition of the corollary is always fulfilled.

Let us show that (\ref{deph}) is the same as the completely
dephasing channel (\ref{phase}). Let
$\mathcal{G}=\{kz_{0}:k=0,1,\ldots ,d-1\}$, then we have
\begin{equation}
\Psi (\rho )=\frac{1}{d}\sum_{k=0}^{d-1}W_{kz_{0}}\rho
W_{kz_{0}}^{\ast }= \frac{1}{d}\sum_{k=0}^{d-1}(W_{z_{0}})^{k}\rho
(W_{z_{0}}^{\ast })^{k}.
\end{equation}
Let
\begin{equation}
W_{z_{0}}|h_{j}\rangle =c_{j}|h_{j}\rangle ;\qquad
c_{j}=\exp\left(\frac{2 \pi {\mathrm i}}{d}\alpha_j \right),
\end{equation}
where all $\alpha _{j}$ must be different $\mathrm{mod}\,d$, for
otherwise two different pure states emerging from the corresponding
eigenvectors would have the same representations (\ref{deco}). Hence
we can assume that ${ \alpha _{j}}=j+{\alpha _{0}};j=0,1,\dots
,d-1$. Therefore we have
\begin{align}
\Psi (|h_{m}\rangle \langle h_{n}|)& =\frac{1}{d}
\sum_{k=0}^{d-1}(W_{z_{0}})^{k}|h_{m}\rangle \langle
h_{n}|(W_{z_{0}}^{\ast
})^{k}  \notag \\
& =\frac{1}{d}\sum_{k=0}^{d-1}\exp \left(\frac{2 \pi {\mathrm
i}}{d}(m-n)k\right)
\vert h_m \rangle\langle h_n \vert  \notag \\
& =
\begin{cases}
|h_{m}\rangle \langle h_{m}| & m=n \\
0 & m\neq n
\end{cases}
.
\end{align} 

Finally consider the case where (\ref{ccom}) is channel. If the
point $ (a,b)\in \mathbf{{R}^{2}}$ is in the triangle, defined by
the corners $(0,1)$, $(-1/(d-1),0)$, $(d/(d-1),-1/(d-1))$ (see
Figure), the function $p_{\gamma } $ is nonnegative for all $\gamma
$ and defines the Weyl-covariant channel (\ref{ch1}). The condition
$\left\vert a+b\right\vert \geq \left\vert b\right\vert $ then
amounts to $a(a+2b)\geq 0$ (this corresponds to the shaded area on
the Figure). In the case of the channel (\ref{ccom}) with $ \Psi $
given by (\ref{phase}) it becomes redundant. The multiplicativity of
$2$-norm in this case follows also from a general result in
\cite{KR2}.
To investigate this case further
in terms of
the additivity of the minimal output entropy and 
the multiplicativity for $p \in [1,+\infty]$,
see \cite{F}.

\section{Complementary channels}

The relation between a channel and its complementary \cite{H2} (conjugate
\cite{KMNR}) was investigated in these papers to show that the
multiplicativity of the original channel implies that of the complementary
channel. Suppose the original channel is given by the Kraus representation
\begin{equation*}
\Phi (\rho )=\sum_{\alpha =1}^{d_{C}}W_{\alpha }\rho W_{\alpha }^{\ast
},\qquad W_{\alpha }:\mathcal{H}_{A}\rightarrow \mathcal{H}_{B}
\end{equation*}
and the complementary channel by
\begin{equation*}
\tilde{\Phi}(\rho )=\sum_{t=1}^{d_{B}}\tilde{W}_{t}\rho \tilde{W}_{t}^{\ast
},\qquad W_{t}:\mathcal{H}_{A}\rightarrow \mathcal{H}_{C}.
\end{equation*}
Here $d_{B}=\mathrm{dim}\mathcal{H}_{B}$ and $d_{C}=\mathrm{dim}\mathcal{H}
_{C}$. Then \cite{H2}
\begin{equation*}
\langle \tilde{e}_{\alpha }|\tilde{W}_{t}=\langle e_{t}|W_{\alpha },
\end{equation*}
where $\{e_{t}\}_{t}$ is an orthonormal basis in $\mathcal{H}_{B}$ and $\{
\tilde{e}_{\alpha }\}_{\alpha }$ in $\mathcal{H}_{C}$.

In this section we compute the complementary of a Weyl-covariant channel.
This was also derived in \cite{KMNR} but we give somewhat more explicit form
by using a different method. In this section we use, for convenience,
different notations for the Weyl-covariant channel
\begin{equation*}
\Phi (\rho )=\sum_{x,y=1}^{d}\lambda _{x,y}^{2}U^{x}V^{y}\rho
(U^{x}V^{y})^{\ast }.
\end{equation*}
Let $e_{t}$ be a row vector with $t$-th entry 1 and others 0. Then
\begin{equation*}
U^{x}=
\begin{pmatrix}
e_{1-x} \\
\vdots \\
e_{t-x} \\
\vdots \\
e_{d-x}
\end{pmatrix}
;\qquad V^{y}=\mathrm{diag}\left[ \exp \left( \frac{2\pi
\mathrm{i}}{d} y\right) ,\ldots ,\exp \left( \frac{2\pi
\mathrm{i}}{d}ty\right) ,\ldots ,1 \right] .
\end{equation*}
Hence
\begin{equation*}
U^{x}V^{y}=
\begin{pmatrix}
\exp \left( \frac{2\pi \mathrm{i}}{d}(1-x)y\right) e_{1-x} \\
\vdots \\
\exp \left( \frac{2\pi \mathrm{i}}{d}(t-x)y\right) e_{t-x} \\
\vdots \\
\exp \left( \frac{2\pi \mathrm{i}}{d}(d-x)y\right) e_{d-x}
\end{pmatrix}
.
\end{equation*}
Therefore reordering the Kraus operators we have
\begin{equation*}
\tilde{W}_{t}=
\begin{pmatrix}
\tilde{W}_{t}^{(1)} \\
\vdots \\
\tilde{W}_{t}^{(s)} \\
\vdots \\
\tilde{W}_{t}^{(d)}
\end{pmatrix}
.
\end{equation*}
Here
\begin{align*}
\tilde{W}_{t}^{(s)}& =
\begin{pmatrix}
\lambda _{d,1-s}\exp \left( \frac{2\pi \mathrm{i}}{d}t(1-s)\right) e_{t} \\
\vdots \\
\lambda _{1-u,1-s}\exp \left( \frac{2\pi \mathrm{i}}{d}(t-1+u)(1-s)\right)
e_{t-1+u} \\
\vdots \\
\lambda _{1,1-s}\exp \left( \frac{2\pi \mathrm{i}}{d}(t-1)(1-s)\right)
e_{t-1} \\
\end{pmatrix}
\\
& =
\begin{pmatrix}
\lambda _{d,1-s}e_{t} \\
\vdots \\
\lambda _{1,1-s}e_{t-1}
\end{pmatrix}
\begin{pmatrix}
\exp \left( \frac{2\pi \mathrm{i}}{d}1\cdot (1-s)\right) & \ldots & 0 \\
\vdots & \ddots & \vdots \\
0 & \ldots & \exp \left( \frac{2\pi \mathrm{i}}{d}d(1-s)\right)
\end{pmatrix}
\\
& =
\begin{pmatrix}
\lambda _{d,1-s} & \ldots & 0 \\
\vdots & \ddots & \vdots \\
0 & \ldots & \lambda _{1,1-s}
\end{pmatrix}
U^{1-t}V^{1-s}  \notag \\
& =D_{s}U^{1-t}V^{1-s},
\end{align*}
where
\begin{equation*}
D_{s}=
\begin{pmatrix}
\lambda _{d,1-s} & \ldots & 0 \\
\vdots & \ddots & \vdots \\
0 & \ldots & \lambda _{1,1-s}
\end{pmatrix}
,
\end{equation*}
which is a diagonal matrix defined for each $s$. Then we have
\begin{equation*}
\tilde{W}_{t}=
\begin{pmatrix}
D_{1} & \ldots & 0 \\
\vdots & \ddots & \vdots \\
0 & \ldots & D_{d}
\end{pmatrix}
\begin{pmatrix}
U^{1-t}V^{d} \\
\vdots \\
U^{1-t}V
\end{pmatrix}
.
\end{equation*}

In general, a channel of the form
\begin{equation*}
\Phi (\rho )=\sum_{k=1}^{N}A_{k}\rho A_{k}^{\ast },
\end{equation*}
can be rewritten as
\begin{equation*}
\Phi(\rho)=
\begin{pmatrix}
A_{1},\ldots ,A_{N}
\end{pmatrix}
(I\otimes \rho )
\begin{pmatrix}
A_{1}^{\ast } \\
\vdots \\
A_{N}^{\ast }
\end{pmatrix}
.
\end{equation*}
Therefore the complementary channel of the Weyl covariant channel can be
written as
\begin{equation*}
\tilde{\Phi}(\rho )=\tilde{W}(I\otimes \rho )\tilde{W}^{\ast },
\end{equation*}
where
\begin{equation*}
\tilde{W}=
\begin{pmatrix}
D_{1} & \ldots & 0 \\
\vdots & \ddots & \vdots \\
0 & \ldots & D_{d}
\end{pmatrix}
\begin{pmatrix}
U^{d}V^{d} & \ldots & U^{1}V^{d} \\
\vdots & \ddots & \vdots \\
U^{d}V^{1} & \ldots & U^{1}V^{1}
\end{pmatrix}
.
\end{equation*}

\bigskip

\textbf{Acknowledgments.} This work was accomplished when A. H. was the Leverhulme
Visiting Professor at DAMTP, CMS, University of Cambridge. The authors are grateful
to Yu. M. Suhov and N. Datta for useful discussions.

\newpage

\begin{center}
\includegraphics[width=12cm, clip]{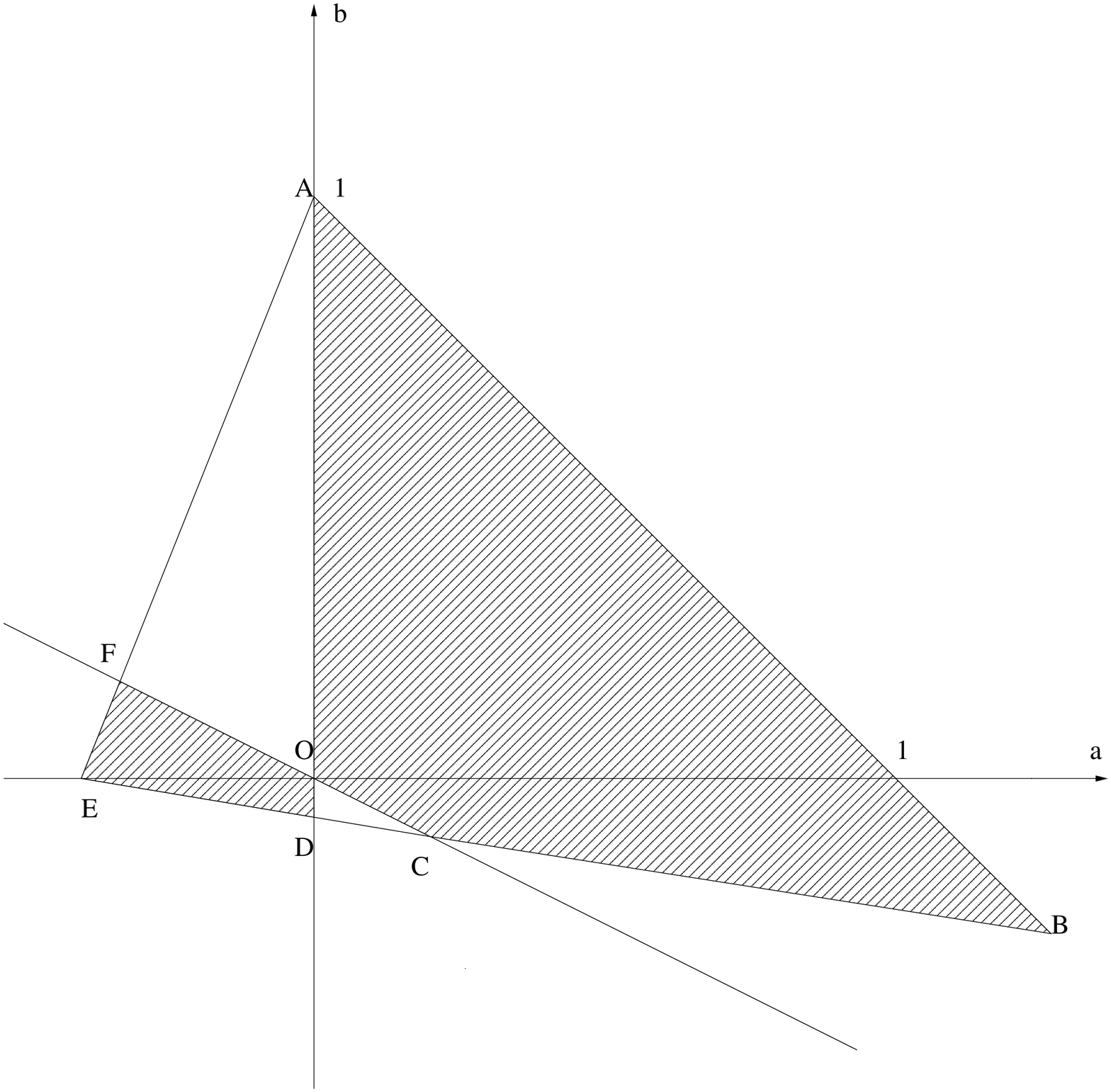} 
\end{center}

\begin{align*}
&\mathrm{A}:(0,1); \qquad
\mathrm{B}:\left(\frac{d}{d-1},-\frac{1}{d-1}
\right);\qquad \mathrm{E}:\left(-\frac{1}{d-1},0\right); \\
&\mathrm{AB}: a+b=1; \qquad \mathrm{AE}:a(d-1)-b=-1; \\
&\mathrm{BE}:a(d-1)+b(d^2-1)=-1; \qquad \mathrm{CF}:a+2b=0
\end{align*}

\end{document}